\documentclass[12pt,english,aps,prl,twocolumn,showpacs,superscriptaddress,groupedaddress,footinbib,reprint,noshowpacs]{revtex4-1}
\usepackage{graphicx}
\usepackage{amsmath}
\usepackage{comment}
\usepackage{amssymb}

\usepackage{subfigure}

\usepackage{subfigure}
\usepackage{color}
\usepackage{soul}

\begin{document}

\title{Optimal design of experiments by combining coarse and fine measurements}
\author{Alpha A. Lee}
\email{aal44@cam.ac.uk}
\affiliation{Cavendish Laboratory, University of Cambridge, Cambridge CB3 0HE, UK}
\affiliation{School of Engineering and Applied Sciences and Kavli Institute of Bionano Science and Technology, Harvard University, Cambridge, MA 02138, USA}

\author{Michael P. Brenner}
\affiliation{School of Engineering and Applied Sciences and Kavli Institute of Bionano Science and Technology, Harvard University, Cambridge, MA 02138, USA}

\author{Lucy J. Colwell}
\email{ljc37@cam.ac.uk}
\affiliation{Department of Chemistry, University of Cambridge, CB2 1EW, Cambridge, UK}

\begin{abstract}
In many contexts it is extremely costly to perform enough high quality experimental measurements to accurately parameterize a predictive quantitative model. However, it is often much easier to carry out large numbers of experiments that indicate whether each sample is above or below a given threshold. Can many such categorical or ``coarse'' measurements be combined with a much smaller number of high resolution or ``fine'' measurements to yield accurate models? Here, we demonstrate an intuitive strategy, inspired by statistical physics, wherein the coarse measurements are used to identify the salient features of the data, while the fine measurements determine the relative importance of these features. A linear model is inferred from the fine measurements, augmented by a quadratic term that captures the correlation structure of the coarse data. We illustrate our strategy by considering the problems of predicting the antimalarial potency and aqueous solubility of small organic molecules from their 2D molecular structure. 
\end{abstract} 

\makeatother
\maketitle

A large class of scientific questions asks whether dependent variables can be accurately predicted by using training data to learn the parameters of quantitative models. Classical statistics shows that this is possible if sufficiently many high resolution measurements are available, though the cost of performing these experiments can be prohibitive. On the other hand, in many settings, it can be straightforward to evaluate whether a  measurement is above or below a certain threshold, raising the question of how such measurements can be incorporated into the modelling framework. 

Examples abound in disparate fields. For instance, predicting the solubility of organic molecules is a fundamental challenge in physical chemistry \cite{llinas2008solubility}. Although accurate measurements are extremely difficult to obtain \cite{palmer2014experimental}, determining whether a molecule is soluble at a particular concentration is comparatively simple. Similarly, in drug discovery, biochemical assays that determine whether a molecule binds to a given receptor are much simpler than measuring protein-ligand binding affinity \cite{malo2006statistical}. In protein biophysics, a key challenge is to predict the effect of amino acid changes on protein phenotype. Here, threshold measurements are naturally provided by homologous sequences from the same protein family  \cite{morcos2014coevolutionary,figliuzzi2015coevolutionary,hopf2015quantification,barrat2016improving,levy2017potts}. In contrast, experimentally measuring the phenotypic change is much more difficult. A related problem is to predict the viral fitness landscape given HIV sequences obtained from patients; again collecting patient samples is much easier than measuring fitness directly \cite{shekhar2013spin,ferguson2013translating}. In single-cell RNA sequencing, decomposition methods that extract the correlation structure of shallow gene expression measurements is an ongoing challenge \cite{dixit2016perturb,buettner2016scalable}. 
	
Despite the ubiquity of this problem, to our knowledge there is no principled method for combining numerous binary/categorical (``coarse'') measurements with fewer quantitative (``fine'') measurements to produce a predictive model. Although regression approaches can account for a prior estimate of sample error \cite{bishop2006pattern}, this is not the same as combining two qualitatively distinct forms of data to build a more accurate model.
	

In this Letter, we introduce an intuitive method that combines coarse and fine measurements. The coarse measurements provide sets of labelled samples -- those data above and below some threshold -- and the proposed method extracts features from the correlations of the variables in each set. Their contribution to the dependent variable is then determined by using the fine measurements to build a regression model for these features. Our model augments a quantitative linear model with a quadratic term which captures the correlation structure extracted from the coarse data. We illustrate our approach by applying it to solubility prediction, and interpret the approach in the context of the Ising model. 

To fix ideas, we assume each sample is characterized by a vector of $p$ properties $\mathbf{f}_i \in \mathbb{R}^p$. The binary data indicates that $N_+$ ($N_-$) samples are above (below) some threshold. In addition, we are given $\mathbf{y} \in \mathbb{R}^{M}$, quantitative measurements for $M$ additional samples. These measurements could be binding affinity, solubility etc. We construct matrices $R_{\pm} \in  \mathbb{R}^{N_\pm \times p}$ for samples above/below the threshold, with columns of $R_{+}$ and $R_{-}$ normalized separately to have zero mean and unit variance.  Intuitively, if there are combinations of the $p$ properties that are always present in either sample set, then these properties should be good predictors of the measurement. Such persistent correlations can be identified from the eigendecomposition of each sample covariance matrix $C_\pm$
\begin{align}
C_{\pm} &= \frac{1}{N_\pm} R_\pm^{T} R_\pm \nonumber\\
&=\sum_{i=1}^{N_\pm} \lambda^{\pm}_i \mathbf{u}^{\pm}_i \otimes \mathbf{u}^{\pm}_i,
\label{sample_eigendecomposition}
\end{align}
where $\{ \lambda^{\pm}_i\}$, $\{ \mathbf{u}^{\pm}_i \}$ are the eigenvalues and eigenvectors (note we perform separate eigendecompositions for the two matrices $C_\pm$).
Each $\mathbf{u}^{\pm}_i$ identifies a particular combination of the $p$ properties, explaining a fraction $\lambda^{\pm}_i/\sum_i \lambda^{\pm}_i$ of the variance \cite{bishop2006pattern}. Each matrix $C_{\pm}$ is an unbiased estimator of the corresponding true covariance matrix. The quality of this estimator depends on data sampling. For example, one may inadvertently assay certain samples (easy to obtain, measure etc.), which could distort the estimator by causing an eigenvector with large eigenvalue to be localized on features common to these samples, even though they do not predict the output variable. For protein sequences, a natural source of such spurious correlations is phylogeny \cite{dutheil2012detecting,obermayer2014inverse}. 

Here we propose that whereas the eigenvectors $\mathbf{u}^{\pm}_i$ reliably identify data features, their significance as estimated by the corresponding eigenvalues $ \lambda^{\pm}_i $ can be severely corrupted by imperfect sampling. Later we justify this ansatz with ideas from statistical physics, and show that this characterization applies to a large class of problems. This ansatz suggests a strategy to mitigate the corruption by using the additional quantitative measurements to determine the significance of each feature. We posit a general quadratic model 
\begin{equation}
y_i = \mathbf{h}^T \mathbf{f}_i + \mathbf{f}_i^{T} J \mathbf{f}_i + \epsilon_i.
\label{regression}
\end{equation}
Here $\mathbf{h}$ is the variable-specific effect, $J$ captures the coupling between variables, and $\epsilon_i \sim N(0,\sigma)$ models random error. There are $p$ parameters in $\mathbf{h}$ and $p(p-1)/2$ parameters in $J$. If one had $M\gg p(p+1)/2$ quantitative measurements, these parameters could be estimated using linear least squares regression. However, it is costly to perform many detailed measurements, so we turn instead to the matrices $C_{\pm}$. We pose the ansatz 
\begin{equation}
J = \sum_{k=1}^{\hat{p}_+} c^{+}_{k}  \mathbf{u}^{+}_{k}  \otimes \mathbf{u}^{+}_{k}  + \sum_{k =1}^{\hat{p}_-} c^{-}_{k}   \mathbf{u}^{-}_{k}  \otimes \mathbf{u}^{-}_{k} .
\label{ansatz}
\end{equation}
Here $\hat{p}_\pm \le p$, since some eigenvectors will reflect noise due to finite sampling \cite{laloux1999noise,plerou1999universal,bai2010spectral}. Our ansatz reflects the hypothesis that the eigendecomposition of $C_{\pm}$ captures variable-variable correlations. If the number of samples is much smaller than the number of variables, random matrix theory provides a rigorous way to determine $\hat{p}_\pm $ \cite{laloux1999noise,plerou1999universal,bai2010spectral,bun2016,lee2016predicting,bun2017cleaning}; this case will be discussed in detail later. Relaxing this assumption, we include all eigenvectors and determine the parameters $\mathbf{h}$, ${c}^{+}$ and ${c}^{-}$ by regressing against the the few quantitative measurements available. We note that the ansatz (\ref{ansatz}) reduces the number of variables to $p+\hat{p}_+ + \hat{p}_- $. In the case where the coarse measurements yield multiple categories (or a single category), our method generalizes by forming separate correlation matrices for each category, and positing that $J$ is a sum of the outer product of all eigenvectors with coefficients determined by regressing against the quantitative data. Our method generalises to Generalised Linear Models with a link function on the right hand side of Equation (\ref{regression}).

To illustrate our approach, we consider two examples: predicting the potency of chemicals against malaria and the equilibrium aqueous solubility of molecules.

\begin{figure}
\centering
\includegraphics[scale=0.18]{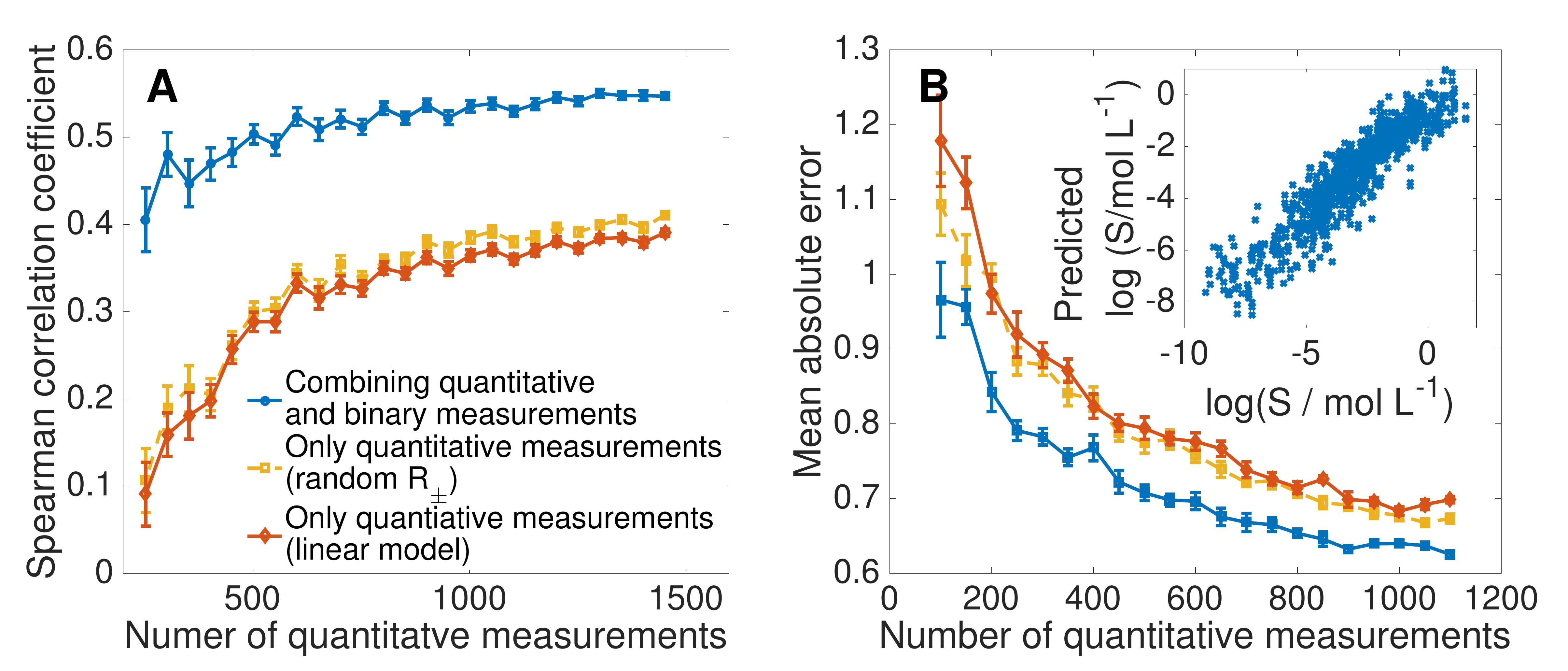}
\caption{Combining coarse and fine measurements accurately predicts antimalarial activity and solubility. The predictive accuracy of (A) $\mathrm{pIC}_{50}$ against malaria and (B) solubility as a function of the number of quantitative measurements given to the model with coarse measurements (blue line), and without (red line). Including random quadratic terms (orange line) is not effective; error bars obtained over 30 random partitions of data into training (90\%) and verification (10\%) sets. (Inset) Out-of-sample solubility prediction with 90\% of the full dataset has a mean absolute error of 0.61 ($r^2$ = 0.85). The estimate is arrived at by analysing 10 random partitions of the data into training and verification sets.}
\label{results}
\end{figure}

\emph{Antimalarials} -- Developing accurate models that can rank the potency of a library of compounds against a target is an important unsolved challenge in drug discovery.  We consider a published antimalarial screening campaign \cite{guiguemde2010chemical}: binary but high throughput assays reported 1528 ÒactiveÓ compounds against malaria, lower throughput but quantitative assays measured the potency ($\mathrm{pIC}_{50}$) of only 1189 compounds \cite{guiguemde2010chemical,Riniker2017}. Figure \ref{results}A shows that by combining binary and quantitative measurements, an hitherto unattempted strategy, an order of magnitude less quantitative measurements could have been performed to yield a model with similar predictive accuracy (c.f. Supplementary Information showing the same result for the Pearson correlation coefficient). Moreover, the model with coarse measurements clearly outperforms the linear model without the coarse measurements and the ``null'' quadratic model where the vectors $\mathbf{u}^{\pm}_k$ are random orthogonal vectors (i.e. random $R_{\pm}$). In the Supplemental Information we show that our model also outperforms direct quadratic regression. The compounds are described using the 1024-bit Morgan6 Fingerprint \cite{rogers2010extended} generated with \texttt{rdKit}~\cite{rdkit}.

\emph{Solubility} -- Predicting the aqueous solubility of molecules is a fundamental problem in physical chemistry important to a plethora of chemical industries. However, accurate solubility assays are low throughput ($\sim$ 1 hour/compound \cite{box2005}). Figure \ref{results}B shows that one could obtain an accurate solubility model ($r^2 = 0.85$, MAE = 0.61) if one were to combine the outcome of a coarse solubility assay that could only tell whether a compound is soluble ($< 10^{-4}~\mathrm{mol/L}$) or not ($> 10^{-2}~\mathrm{mol/L}$) with much fewer quantitive solubility data. We use a standard dataset of the solubility of 1144 organic molecules \cite{delaney2004esol}, and describe the molecule by concatenating the Avalon Fingerprint \cite{gedeck2006qsar}, the MACCS Fingerprint \cite{durant2002reoptimization}, and the 1024-bit Morgan6 Fingerprint \cite{rogers2010extended}.  Our result compares favourably with other models that also use binary molecular fingerprints, e.g. kernel partial least squares regression achieves $r^2 = 0.83$ \cite{zhou2008scores}. 

To understand why our heuristic strategy is successful, we consider a model problem where data is generated according to Eqn. (\ref{regression}), which is the maximum entropy model \cite{schneidman2006weak,cocco2009neuronal,lee2015statistical}, analogous to the Ising model. We thus interpret the dependent variable as an ``energy'', noting that the logarithm of the solubility is proportional to the solvation energy. The interaction matrix $J$ can be decomposed into a sum of outer products of eigenvectors $\boldsymbol{\zeta}_i $ (Hopfield patterns \cite{hopfield1982neural}), and eigenvalues $E_i$ (Hopfield energies) as
\begin{equation} 
J = \sum_{i=1}^{m} E_i \boldsymbol{\zeta}_i  \otimes \boldsymbol{\zeta}_i. 
\end{equation}
Furthermore, to model the binary features used in solubility prediction,  we make the assumption that the independent variable is a vector of $\pm 1$. 

To simulate binary measurements we randomly draw samples from the uniform distribution, evaluate Eqn. (\ref{regression}) to determine the energy of each sample, and retain those samples that fall below a certain energy. Consider an interaction matrix $J$ with $p=100$, and $m=3$ randomly generated patterns. To fix ideas, henceforth let all patterns be attractive with $(E_1,E_2,E_3) = (- 30, - 25, - 20)$, $\mathbf{h}=0$ and $\epsilon_i=0$. Using this model, we generate 5000 random vectors, consider the $N = 500$ samples with lowest energy as above threshold samples, and compute the eigendecomposition of the resulting correlation matrix. 

\begin{figure}
\centering
\includegraphics[scale=0.30]{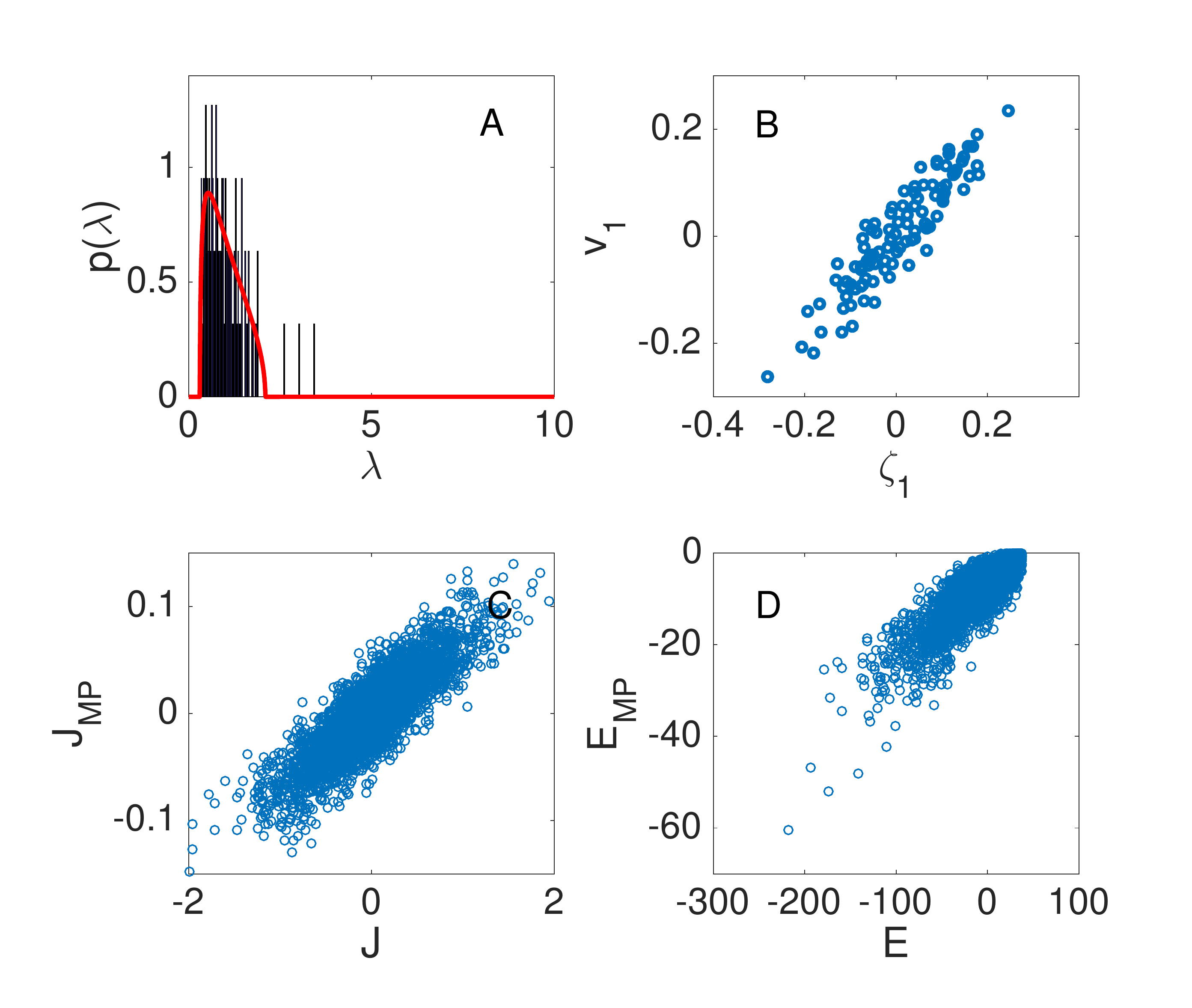}
\caption{Hopfield patterns can be recovered from threshold sampling. (A) Histogram of eigenvalues agrees with the Mar\v{c}enko-Pastur distribution (red curve) save for three significant eigenvalues. (B) The top eigenvector is the lowest energy Hopfield pattern; the other eigenvectors are shown in Supplemental Information. Random matrix cleaning allows us to successfully (C) recover the coupling matrix $J$ and (D) predict Hopfield energies. }
\label{pattern}
\end{figure}

Figure \ref{pattern}A shows that the eigenvalue distribution of the sample correlation matrix $C_{+}$ follows the Mar\v{c}enko-Pastur distribution expected for a random matrix \cite{marvcenko1967distribution}, 
\begin{equation} 
\rho(\lambda) =  \frac{\sqrt{ \left[ \left(1+\sqrt{\gamma} \right)^2 - \lambda \right]_+ \left[  \lambda - \left(1-\sqrt{\gamma} \right)^2 ) \right]_+}}{2 \pi \gamma \lambda} 
\label{MP}
\end{equation} 
where $(\cdot)_+ =\mathrm{max}(\cdot,0)$, $\gamma = p/N$, with the exception of three distinct eigenvalues. Figure \ref{pattern}B shows that their corresponding eigenvectors are indeed the Hopfield patterns that we put in. Therefore, the large eigenvectors of $C_{+}$ correspond to eigenvectors of $J$. Note that the random matrix theory framework applies because $m\ll p$, i.e. the signal is low rank compared to the noise. If the signal was high rank, all eigenvectors should be included and their significance determined by regression against fine measurements, as in the examples discussed above.

Turning to eigenvalues, in this model, which features \emph{uniform} sampling, we find that the eigenvalues are proportional to the Hopfield energy $E_i$. This allows us to ``clean'' the correlation matrix, by using the $q$ eigenvectors above the Mar\v{c}enko-Pastur threshold to construct a rank $q$ approximation $J_{\mathrm{MP}}$ of the correlation matrix (here $q = 3$).
Figure \ref{pattern}C shows that $J_{\mathrm{MP}}$ accurately reconstructs $J$, and allows accurate prediction of the energy of particular states (Figure \ref{pattern}D). Analogously, the eigendecomposition of $C_-$ allows one to recover \emph{repulsive} patterns with positive Hopfield energies (see Supplemental Information). 



Since the Hopfield patterns are energy minima, taken together Figure \ref{pattern}A-D imply that the probability of the system visiting a particular basin under uniform sampling is proportional to the energy of that minima. Therefore, the hypervolume of each energy basin is proportional to the basin depth. This fact can be derived by noting that Eqn. (\ref{regression}) is a quadratic form, so the Hessian matrix is a constant. Therefore, all local minima have the same mean curvature.  Given that low lying energy minima are wide, we can extract the position of energy minima in the space of input variables by studying the correlation structure of the binary dataset. We note that the correlation between basin hypervolume and basin depth appears in many complex physical systems beyond the Ising model \cite{doye1998thermodynamics,doye2005characterizing,pickard2011ab}. 

\begin{figure}
\centering
\includegraphics[scale=0.24]{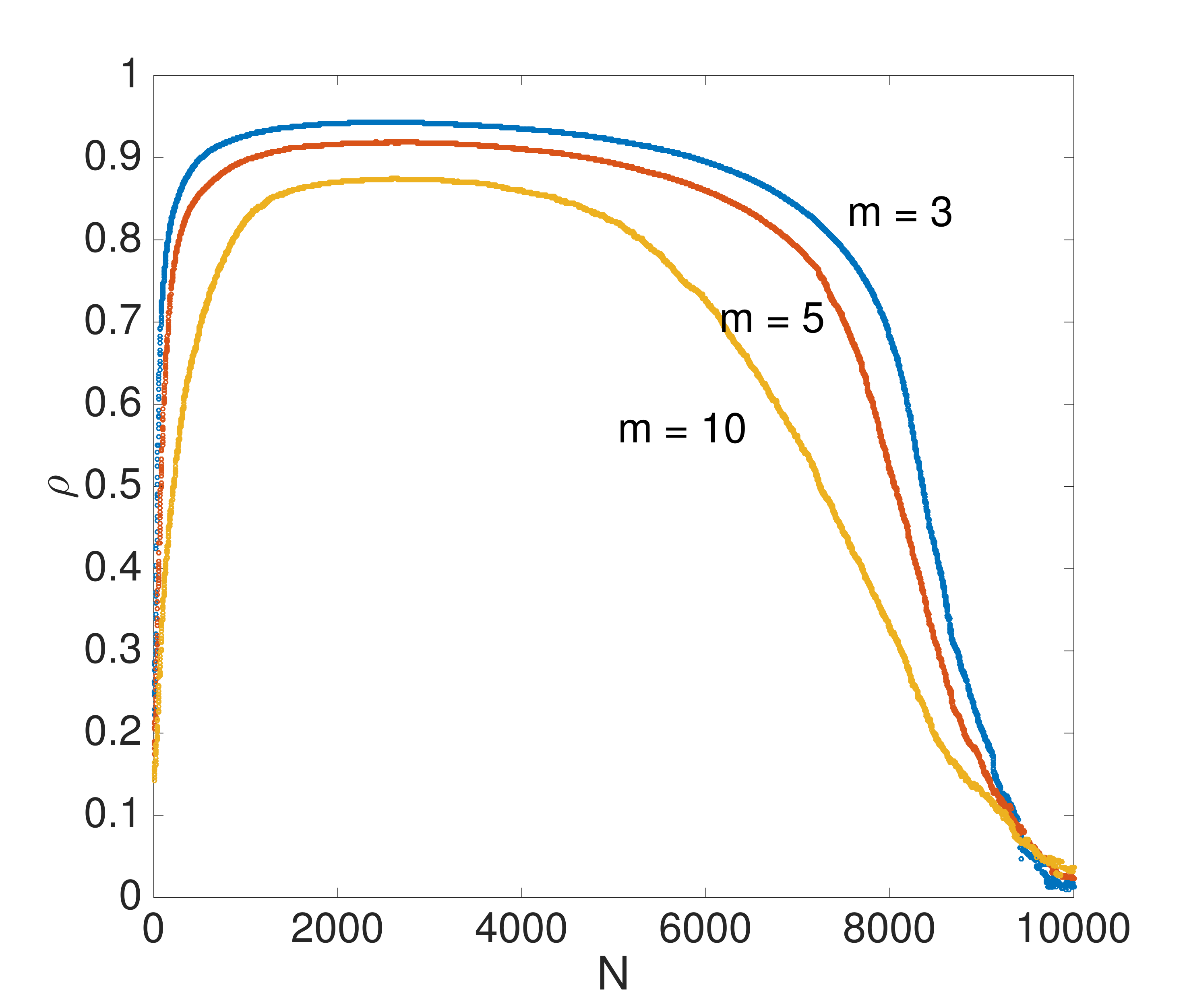}
\caption{Hopfield inference with random matrix cleaning is robust to the energy threshold. Here $\rho$ is the Pearson correlation coefficient between the entries in $J$ and $J_{\mathrm{MP}}$. The results are computed by averaging over $50$ realizations. }
\label{sample} 
\end{figure}

A lingering question is whether our inference procedure is robust to the choice of threshold. To test this, we consider $m$ Hopfield patterns, chosen as eigenvectors of a symmetrized $p\times p$ Gaussian random matrix, with the Hopfield energy chosen to be Gaussian distributed with mean $10$ and unit variance. We draw 10000 samples randomly and compute the correlation matrix with the lowest energy $N$ samples. Figure \ref{sample} shows that our method is robust: the correlation coefficient between $J$ and $J_{\mathrm{MP}}$ is large and constant for a wide range of thresholds and number of Hopfield patterns. The question of how many energy minima can be recovered from the binary data and a thermodynamic interpretation is discussed in the Supplemental Information. 

We now turn to consider two common scenarios that break the assumptions made so far, stratified sampling and more complex energy landscapes. 

\begin{figure}
\centering
\includegraphics[scale=0.33]{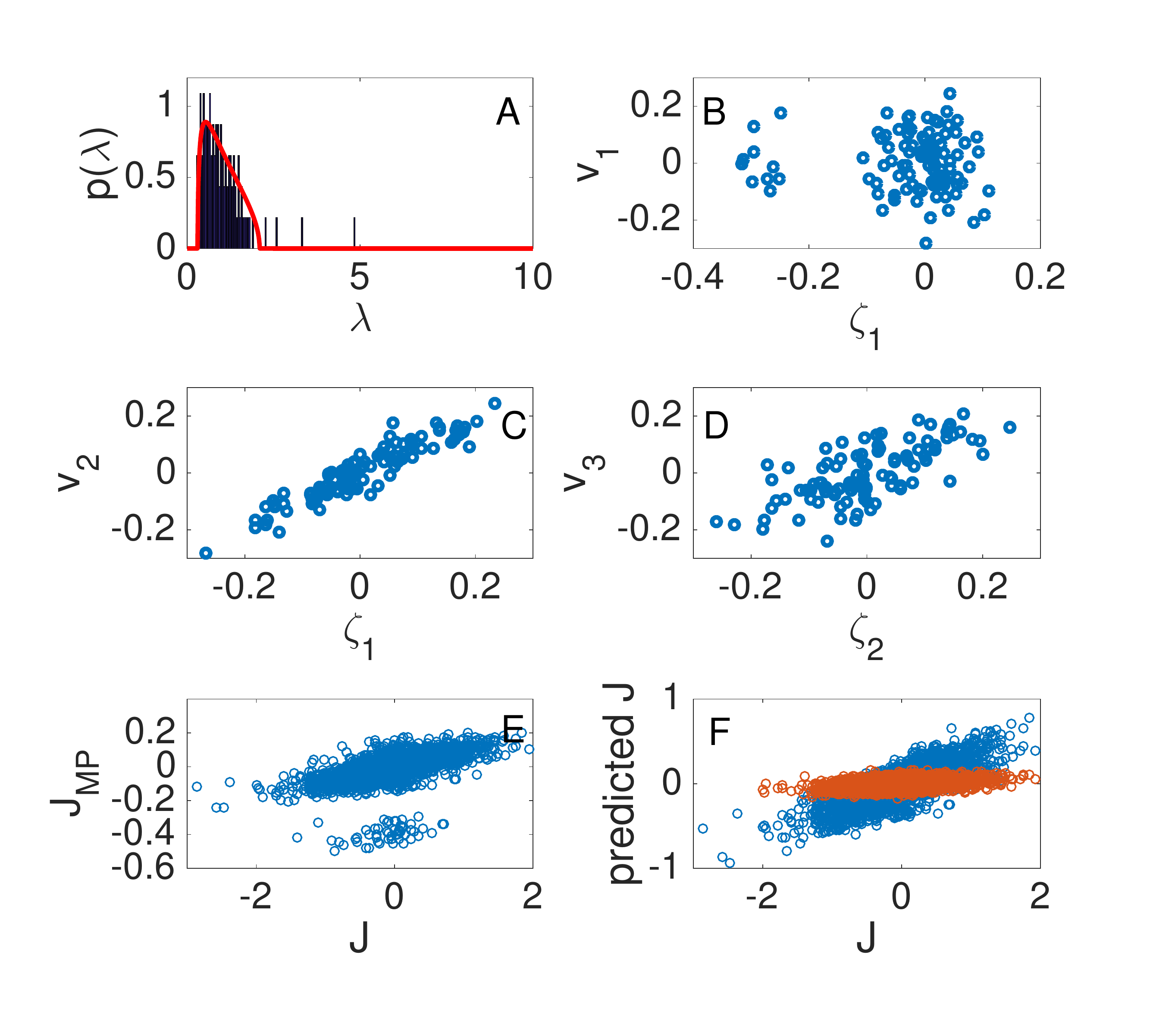}
\caption{Few quantitative measurements enable $J$ to be inferred accurately for stratified datasets. (A) There are now four significant eigenvectors but still only three Hopfield patterns in the model. (B)-(D) The top eigenvector is uncorrelated with all Hopfield patterns, and the Hopfield patterns are demoted to the second to fourth significant eigenvectors. (E) Random matrix cleaning does not recover the coupling matrix. (F) Blue data points: The elements of the coupling matrix recovered by incorporating quantitative data and using Eqn. (\ref{ansatz}); Orange data points: the elements of the coupling matrix recovered using the quantitative data only and ridge regression (with $J_{ij}$ being the coefficients).  }
\label{freeze_spin}
\end{figure}

\emph{Stratified sampling}: Thus far we assumed that the sampling is uniform before thresholding. However, in many settings, the sampling is biased. To model this effect, we draw 5000 random samples, but freeze the first 5 variables to $+1$ for the first 2500 samples and the last 5 variables to $-1$ for the remaining 2500 samples. We then evaluate the energy, and take again the lowest $10^{\mathrm{th}}$ percentile. Figure \ref{freeze_spin}A-D shows that the frozen variables introduce sample-sample correlations, and now there are 4 significant eigenvectors with the first Hopfield pattern demoted to the second largest eigenvector (Figure \ref{freeze_spin}C). As such, the informative eigenvectors are still present in $J_{\mathrm{MP}}$, but the eigenvalues are misplaced. 

In this case, na\"{i}ve random matrix cleaning does not recover $J$ (Figure \ref{freeze_spin}E), since there is no \emph{a priori} reason to discard the first eigenvector unless we know the Hopfield patterns beforehand. We need additional information -- for which we turn to the quantitative measurements -- to accurately recover the Hopfield energies. Figure \ref{freeze_spin}F shows that an additional 500 quantitative measurement allow us to recover the coupling matrix (MAE = 0.12) using ridge regression and the ansatz Eqn. (\ref{ansatz}). The error is significantly larger (MAE = 0.31) if only the quantitative measurements are used. 

\emph{Complex energy landscapes}: The geometric property that the depth of an energy minima is related to its hypervolume is not universal to all energy landscapes \cite{wales1998archetypal,wales2003energy}. A natural question is whether the significant eigenvectors and eigenvalues of the correlation matrices of samples below/above an energy threshold allow us to infer features of a complex energy landscapes. We consider a landscape that comprises a sum of Gaussians
\begin{equation}
H(\mathbf{f}) = \sum_{i} E_i \exp\left( - E_i^2 (\mathbf{f}\cdot \boldsymbol{\zeta}_i)^2\right). 
\label{gaussian_landscape}
\end{equation} 
This landscape has the property that the depth of each energy minima, $E_i$ (located at $\boldsymbol{\zeta}_i$), is \emph{inversely} proportional to its width $1/E_i$. As above, we let $(E_1,E_2,E_3) = (- 30,- 25, - 20)$ and generate Hopfield patterns by diagonalising a symmetrized Gaussian random matrix. We draw $5000$ samples and threshold to find the $500$ lowest energy samples. Figure \ref{gaussian_landscape} shows that there are again three significant eigenvectors above the Mar\v{c}enko-Pastur threshold, but the lowest energy Hopfield pattern is demoted to the third eigenvector, while the highest energy Hopfield pattern is promoted to the top eigenvector. This is expected: the eigenvalue corresponding to each minimum is proportional to the number of samples near that minimum, i.e. the basin volume, which in this case is not proportional to basin depth. However, the eigenvectors still indicate the locations of the energy minima, motivating the approach described in Eqn. \eqref{regression}, where we use these eigenvectors to identify features. 


\begin{figure}[h]
\centering
\includegraphics[scale=0.38]{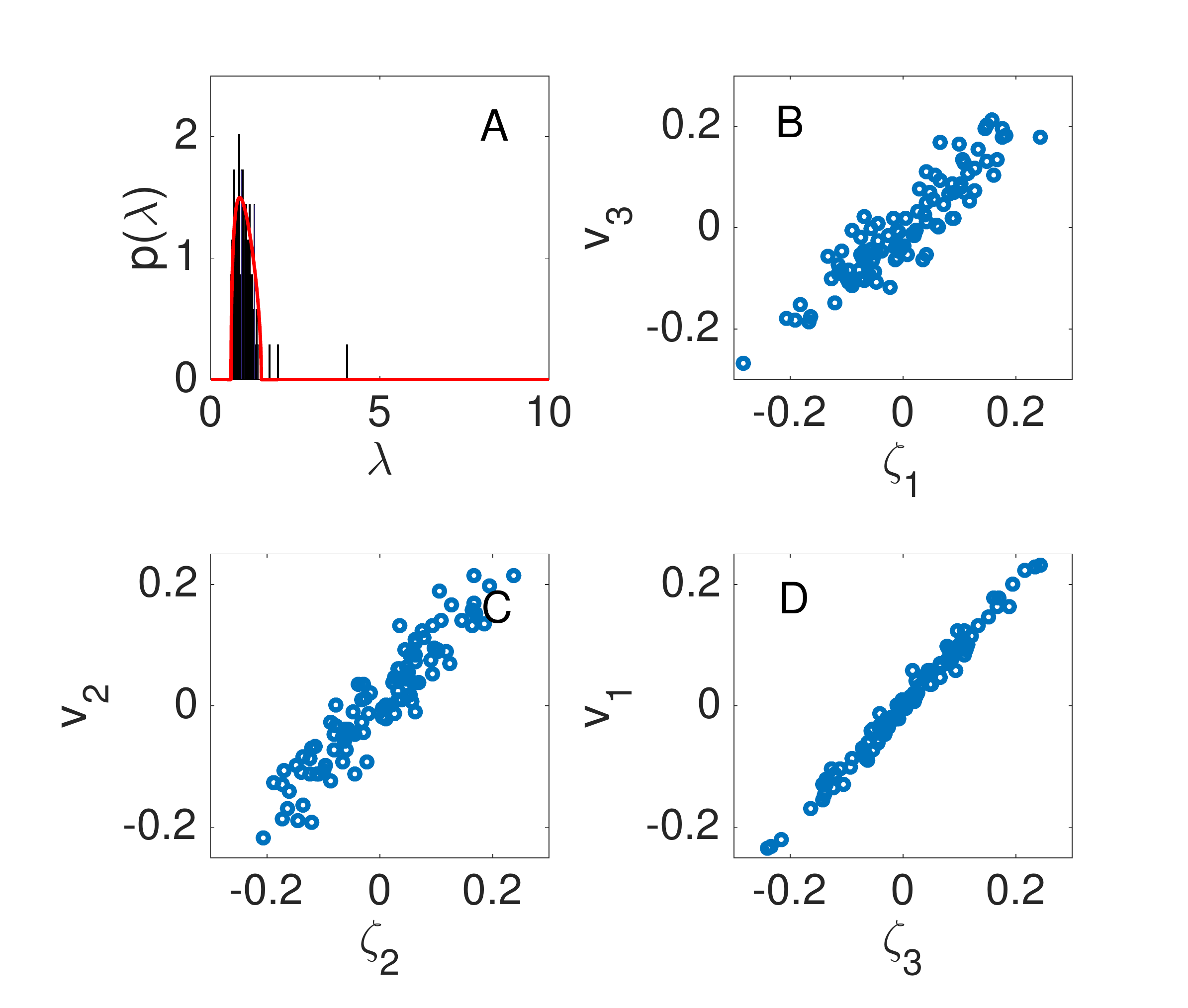}
\caption{For energy landscapes where basin depth is not proportional to basin width, the eigenvectors indicate the locations of energy minima but the eigenvalues are awry. (A) There are three significant eigenvectors above the  Mar\v{c}enko-Pastur threshold. (B) - (D) The top eigenvector is correlated with the highest energy minimum, and the last significant eigenvector is correlated with the lowest energy minimum.}
\label{gaussian_landscape}
\end{figure}

In conclusion, we develop a general strategy, grounded in statistical physics, which integrates coarse and fine measurements to yield a predictive model. Since coarse measurements are often significantly less costly to obtain, our strategy provides a new avenue for experiment design. Although our Letter only considered an Ising-type model, the fact that the eigenvectors of the correlation matrix of coarse measurements point toward energy minima suggests a natural way to integrate our result into more complex non-linear models, for example by using $\mathbf{f}\cdot \mathbf{u}_i$, the overlaps between the sample vector and each eigenvector, as inputs to a general nonlinear function such as an artificial neural network.

\acknowledgments
The authors thank R Monasson for insightful discussions. AAL acknowledges the support of the George F. Carrier Fellowship and the Winton Advanced Research Fellowship.
LJC acknowledges a Next Generation fellowship, and a Marie Curie CIG [Evo-Couplings, Grant 631609]. MPB is an investigator of the Simons Foundation, and acknowledges support from the National Science Foundation through DMS-1411694.

\bibliography{hopfield_refs} 
\end{document}